\documentclass{cpcauth}

\usepackage{graphicx,epsf}
\usepackage[figuresright]{rotating}

\journal{cpc}

\begin{document}

\begin{frontmatter}

\title{Numerical Implementation of Three-Body Forces in 
Bound State Faddeev Calculations
in Three Dimensions}

\author[athens]{H. Liu}
\author[athens,fzj]{Ch. Elster \thanksref{lab1}}
\thanks[lab1]{Corresponding author}
\ead{elster@ohiou.edu}
\author[bochum]{W. Gl\"ockle}

\address[athens]{Institute for Nuclear and Particle Physics, Ohio University, Athens OH 45701}
\address[fzj]{Insitut f\"ur Kernphysik, Forschungszentrum J\"ulich, D-52425 J\"ulich}
\address[bochum]{Institut f\"ur Theoretische Physik II, Ruhr-Universit\"at
  Bochum, D-44780 Bochum}

\begin{abstract} 
The Faddeev equations for the three-body bound state are solved directly as three-dimensional
integral equations without employing partial wave decomposition. Two-body forces 
of the Malfliet-Tjon type  and simple spin independent genuine three-body forces
are considered for the calculation of the three-body
binding energy. 
\end{abstract}

\vspace{-4mm}

\end{frontmatter}

\maketitle

\section{Introduction}
 
Three nucleon (3N) bound state calculations are traditionally carried out
by solving Faddeev equations on a partial wave basis. After truncation
this leads to a set of a finite number of coupled equations in two
variables for the amplitude.  The calculations are performed either in momentum
space \cite{pickle,statler,nbench}, in configuration space
\cite{payne,schell0}, or in a hybrid fashion using both spaces
\cite{wu}. Though a few partial waves often provide qualitative
insight, modern three nucleon bound state calculations need 34 or more
different isospin, spin and orbital angular momentum combinations
\cite{nbench}.  It appears therefore natural to avoid a partial wave
representation completely and work directly with vector variables \cite{trit3d}.
This is especially the case, if one wants to consider genuine three-nucleon force (3NF)
effects. The true advantage stems from the fact that a 3N calculation is carried out in Jacobi
variables, whereas a typical 3NF has the form of two consecutive meson-exchange propagators
between e.g. particles 1 and 2, then particles 2 and 3. Using vector variables, the necessary
coordinate transformations are numerically realized by interpolations, whereas in a partial
wave based calculation a large amount of coupling coefficients has to be evaluated and
stored~\cite{hueber2}, requiring large storage and memory capabilities of a computer
architecture. 

\section{Bound State Equation  with a Three-Body Force} 
The Faddeev component describing the bound state of three identical particles interacting via
pairwise forces as well as genuine three-body forces can be written as 
\begin{equation}
\psi_i = G_0 t_i P \psi_i + (1 + G_0 t_i) G_0 W^{(i)}(1+P) \psi_i , \label{eq:1}
\end{equation}
where $i=1,2,3$. In the following, we choose $i=1$ without loss of generality.
The operator $t_i$ stands for the two-body
t-matrix in the subsystem $(jk)$ summing up the pair interaction in this system.  The
quantity $W^{(i)}$,
shown diagrammatically in Fig.~1, is defined via
\begin{equation}
V_{123} = W^{(1)} +W^{(2)} +W^{(3)},  \label{eq:2}
\end{equation}
where $ W^{(i)}$ is that part of the full 3N force $V_{123}$ which is
symmetric for the exchange of nucleons $j$ and $k$ ($j\neq i\neq k$).
The decomposition of Eq.~(\ref{eq:2}) is natural, e.g. for the
$\pi \pi$ exchange 3N force which is present in all currently available
3N forces. The free 3N propagator is given by 
\begin{equation}
G_0^{-1}=E-\frac{p^2}{m}-\frac{3}{4m}q^2,
\end{equation}
 where
{\bf p} and {\bf q} are the standard Jacobi momenta
\begin{eqnarray}
{\bf p}_i &=& \textstyle{\frac{1}{2}}({\bf k}_j - {\bf k}_k)  \\ \label{eq:3} \nonumber
{\bf q}_i &=& \textstyle{\frac{2}{3}}({\bf k}_i - \textstyle{\frac{1}{2}}
({\bf k}_j+{\bf k}_k)),
\end{eqnarray} 
where $ijk=123$ and cyclic permutations thereof. 
The permutation  operator $P$ is given as $P= P_{12}P_{23}+P_{13}P_{23}$, and
the full 3N wave function is related to the Faddeev component by
\begin{equation}
|\Psi \rangle = (1+P) |\psi \rangle. \label{eq:3a}
\end{equation}
\begin{figure}
\begin{center}
\includegraphics*[width=1.8cm]{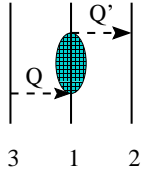}
\end{center}
\caption{Two meson exchange three-nucleon force}
\label{fig:3nf}
\end{figure}
The solution of Eq.~(\ref{eq:1}) with $W^{(1)}=0$ is described in detail in Ref.~\cite{trit3d}
and shall not be discussed here. For our model calculations we also use  Yukawa
interactions of the Malfliet-Tjon type here, 
however, we modify the interaction with a cutoff function of
a dipole type \cite{nonlocal}.

In order to develop the algorithm, we concentrate on a model 3N force of scalar meson exchange
character, which can be written as 
\begin{equation}
W^{(1)} = \frac{1}{(2\pi)^6}  \frac{a_s}{m_s} g_s^2  \frac{F(Q^2)}{Q^2+m_s^2}  \:
        \frac{F(Q'^2)}{Q'^2+m_s^2},   \label{eq:5}
\end{equation}
where $m_s$ represents the mass of the exchanged meson. The momenta ${\bf Q}$ and ${\bf
Q'}$ are defined in Fig.~\ref{fig:3nf} and given by
\begin{eqnarray}
{\bf Q} & =&  {\bf p} -{\bf p}' -\frac{1}{2}({\bf q} - {\bf q}')  \nonumber \\
{\bf Q}'&  =&  {\bf p} -{\bf p}' +\frac{1}{2}({\bf q} - {\bf q}'). \label{eq:6}
\end{eqnarray}
The function $F(Q^2)=\left[ (\Lambda_s^2-m_s^2)/(\Lambda_s^2+Q^2) \right]^2$ represents a
cutoff function for large momenta $Q^2$. 
The parameters of the 3NF used in the presented calculation are given in
Table~1.
\begin{table}
\begin{center}
\begin{tabular}{cccc}
\hline \hline
  $g^2_s/4\pi$ & $m_s$(MeV) & $\Lambda_s$(MeV) &
  $a_s$\\ \hline
 5.0 & 305.8593 & 1000.0 & -1.73  \\ \hline \hline
\end{tabular}
\end{center}
\caption{The parameters of the three-nucleon force.}
\end{table}
For evaluating Eq.~(\ref{eq:1}) we need to calculate matrix elements like
\begin{equation}
\langle {\bf p}{\bf q}| G_0 W^{(1)} | \Psi \rangle.  \label{eq:7}
\end{equation}
Considering the momentum dependence in the meson propagators of $W^{(1)}$, one can easily
imagine that the efficiency of any algorithm will crucially depend on the choice of coordinate
system(s) when carrying out the integrations over the intermediate momenta.
Our numerical evaluation of Eq.~(\ref{eq:7}) is based  on the realization that 
$W^{(1)}$ can be interpreted as two independent interactions, first in the subsystem $(12)$,
then in $(31)$. Explicitly, we write 
\begin{eqnarray}
\langle {\bf p}{\bf q} & &| W^{(1)} | \Psi \rangle = \int d^3 p' \int d^3 q' 
                  \  _1\langle {\bf p}{\bf q} | {\bf p'}{\bf q'}  \rangle_2 \nonumber \\
 & &  \int d^3 p'' \frac{F(({\bf p'} -{\bf p''})^2)}{({\bf p'} -{\bf p''})^2+m^2_s}
   \nonumber \\
& & \int d^3 p''' \int d^3 q'''  \ _2\langle {\bf p''}{\bf q'} |{\bf p'''}{\bf q'''}  \rangle_3
 \nonumber \\
& & \int d^3 p'''' \frac{F(({\bf p'''} -{\bf p''''})^2)}{({\bf p'''} -{\bf p''''})^2+m^2_s}
 \ _3 \langle {\bf p''''}{\bf q'''} | \Psi  \rangle .
\end{eqnarray}
Here the subscripts $1,2,3$ of the bra and ket vectors define the meaning of the related
vectors ${\bf p}$ and ${\bf q}$,  namely the particle number $i$ singled out by ${\bf q_i}$ 
as indicated in Eq.~(4).
We would like to point out that in each integration over a piece of
the 3NF we only integrate over three variables, namely the magnitude of a momentum and two
angle variables. It turns out that it is most favorable to choose momenta ${\bf q'}$ 
(${\bf q'''}$) parallel to the z-axis, so that no interpolation on the unknown function
$\Psi (p,q',{\hat p}\cdot {\hat q'})$ needs to be performed. The calculation of the 
transformation from one subsystem to another, e.g. $_2\langle {\bf p''}{\bf q'} |{\bf
p'''}{\bf q'''}  \rangle_3 $, requires  three-dimensional interpolations. We employ cubic
Hermite splines \cite{hueber2}, which prove to be both accurate and fast.

\begin{table}
\begin{center}
\begin{tabular}{cccccc}
\hline \hline
 $g^2_a/4\pi$ & $m_a$(MeV) & $\Lambda_a$ (MeV) & $g^2_{r}/4\pi$ &
$m_{r}$(MeV) & $\Lambda_r$(MeV)\\ \hline
   -3.5775 & 330.2104 & 1500 & 9.4086 & 612.4801 & 1500 \\ \hline \hline
\end{tabular}
\end{center}
\caption{The parameters of the two nucleon pair force.}
\end{table}

\section{Computational Approach}
The discretized Faddeev equation for a bound state (neglecting spin degrees of freedom) is an
integral equation in  three variables on a typical grid size of 65*65*42 (momentum magnitudes, $p$,
$q$, and angle between the momentum vectors). 
A priori the multidimensionality of the integral
equation to be solved requires more memory. However, on an MPP system the number of variables
and thus the memory do not pose a computational problem, since a variable defining 
a specific 
dimension of the grid can be distributed over a number of processors, leaving a lower
dimensional grid on each processor. As such, our three-dimensional approach is ideally suited
as MPP application, and we can achieve an almost perfect load balance in our runs.

\begin{figure}
\begin{center}
\includegraphics*[width=7.6cm]{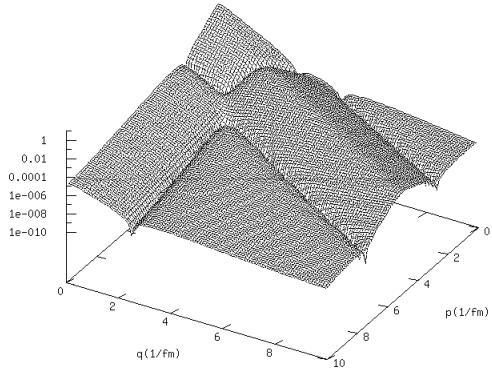}
\end{center}
\caption{The magnitude (in fm$^3$) of the
3N bound state wave function $\Psi(p,q,x))$ for $x=1$ calculated
with the potential parameters listed in Tables 1 and 2.}
\label{fig:totalw}
\end{figure}

The eigenvalue equation for the bound state is
solved iteratively by using Lanczo's type  techniques, here the method of iterated orthogonal
eigenvectors \cite{statler}. For a typical run ten orthogonal eigenvectors are calculated per
energy. We need about 5 to 7 energy iterations to find the ground state energy.

The calculation of the kernel of the integral equation means evaluating the matrix elements
given on the right-hand side of Eq.~(\ref{eq:1}) on a fixed grid $p$, $q$, and angle $x=\cos
({\hat p}\cdot {\hat q})$. The two-body t-matrix (with the two-nucleon pair interaction as
the driving term) 
is obtained by solving a system of linear equations of the form $A \cdot x = b$,
where A is typically a 2500*2500 matrix. This system is solved for about 60 different vectors
$b$, distributed over correspondingly many processors. The integrations over the 3NF terms are
also distributed over the same number of processors, which depends on the size of the q-grid.
Details about the different grid sizes and the dependence of the numerical accuracy on them
can be found in Ref.~\cite{trit3d}.

\section{Results and Discussion}

For our model calculation we use Yukawa  interactions of the Malfliet-Tjon \cite{MT} type
\begin{eqnarray}
V ({\bf p'},{\bf p})& =&  \nonumber \\
\sum_{\alpha=a,r} & &  \frac{g^2_{\alpha}}{({\bf p'}-{\bf p})^2+
m^2_\alpha }  \left( \frac {\Lambda_\alpha ^2 - m^2_\alpha }{\Lambda_\alpha ^2 + 
({\bf p'}-{\bf p})^2} \right)^2  \label{eq:8}
\end{eqnarray}
with a short-ranged repulsive and a long-ranged attractive piece. The parameters used in our
calculation are given in Table~2.
A calculation of the three-body binding energy with this pair force gives $E_t$=7.699~MeV. The
parameters of the 3NF are then adjusted such that its inclusion in the calculation gives a
binding energy $E_t$=8.590~MeV, a value close to the measured one. 
The full 3N wave function is calculated from the solution of  Eq.~(\ref{eq:1}) using
 Eq.~(\ref{eq:3a}) and shown in Fig.~2 
as a function of the momenta $p$ and $q$ at the fixed angle $x=1$. 
A coparison of the 3N wave functions calculated with the 2N pair forces alone and with
the inclusion of the 3NF is shown in Figs. 3 and 4. Both figures depict slices
through the wave functions at specific values.

\begin{figure}
\begin{center}
\includegraphics*[width=4.0cm,angle=-90]{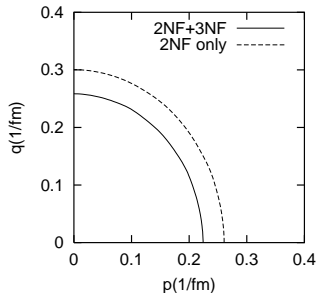}
\end{center}
\caption{Contour slices of the 3N wave function at 2.4 fm$^3$. The solid line represents the
wave function calculated with 2N pair forces and 3N forces, the dashed line stands for the
wave function calculated with 2N pair forces only.}
\end{figure}

\begin{figure}
\begin{center}
\includegraphics*[width=4.0cm,angle=-90]{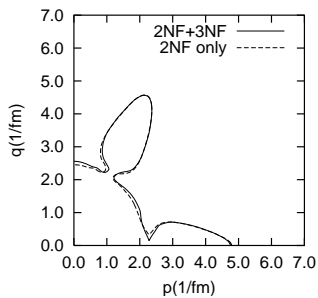}
\end{center}
\caption{Same as Fig.~3, except that the 
contour slices of the 3N wave function are taken at 10$^{-3}$fm$^3$.}
\end{figure}

The 3NF in our model calculation is attractive, as expressed by the larger binding
energy. One would expect that the system becomes slightly smaller and 
acquires more high momentum components. Since the wave functions are both normalized to one,
an increase  of high momentum components will be seen as a decrease in low momentum
components, as seen in Fig.~3. Once the wave function decreases by some orders of magnitude, the
differences disappear.

In summary, 
an alternative approach to state-of-the-art three-nucleon bound state
calculations, which are based on solving the Faddeev equations on a partial wave basis,
is to work directly with momentum vectors. 
We formulate the Faddeev equations for identical
particles as a 
function of vector Jacobi momenta, specifically the magnitudes of the momenta and
the angle between them, for the case when not only pair forces act between the particles
but also genuine three-body forces. As model forces we concentrate on scalar forces, a
superposition of an attractive and repulsive Yukawa interaction for the pair force and an
attractive Yukawa interaction for the three-body force. We demonstrate the numerical
feasibility and accuracy of the solution. We want to point out that the incorporation of the
3N forces is less cumbersome in a three-dimensional approach,  and the algorithm can be
made quite efficient on parallel architectures.

\vspace{1mm}

{\bf Acknowledgments} This work was performed in part under the auspices of 
the U.~S.  Department of Energy under contract
No. DE-FG02-93ER40756 with Ohio University.
We thank the the 
Neumann Institute for Computing (NIC)
and the National Energy Research Supercomputer Center (NERSC) for the use of
their Cray-T3E computers.

\end{document}